\documentclass[twoside,11pt]{article}
\def\be{\begin{equation}}
\def\ee{\end{equation}}
\def\ba{\begin{align}}
\def\ea{\end{align}}
\def\bea{\begin{eqnarray}}
\def\eea{\end{eqnarray}}

\def\ba{\bar{\mathcal A}}

\usepackage{amsmath,amssymb,amsfonts}
\usepackage{amsfonts}
\usepackage{graphicx}
\usepackage{graphics}
\usepackage{breqn}
\begin{document}
\setlength{\unitlength}{10mm}

\title{A Gauged Open 2-brane String in the p-brane Background}

\author{
Fahimeh Sarvi
\\
Majid Monemzadeh
\\
Salman Abarghouei Nejad
\\
{\scriptsize Department of Physics,
}\\{\scriptsize University of Kashan, Kashan 87317-51167, I. R. Iran}}

\date{}
\maketitle
\vspace{1cm}
\vspace{-1.5cm}\begin{abstract}
\noindent
In this article, we make a gauge theory from the Open p-brane system and map it into the Open 2-brane one. Due to the presence of second class constraints in this model, we encounter some problems during the procedure of  quantization. In this regard, considering boundary conditions as Dirac conditions, one can drive the constrained structure of the model at first. Then, with the help of BFT formalism of constraint systems, the Open 2-brane model is embedded into an extended phase space. For this purpose, we introduce some tensor fields to convert ungauged theory into the gauged one. This is the novel part of our research, while mostly scalar and vector fields are used to convert second class constraints into first ones.


\end{abstract}
\textbf{keywords :}Gauge Theories, Constrained Systems, Second Class Constraints, Open p-brane, BFT formalism.
\newpage

\newcommand{\f}{\frac}

\newtheorem{theorem}{Theorem}[section]
\newcommand{\sta}{\stackrel}
\section{Introduction}
In the scenario of point-like particles, the infinite mass leads to several infinities, which obliged physicists to developed methods of renormalization to overcome these problems. Afterwards, although string theory as a theoretical frame work provided some tools to simplify the unification of general relativity and quantum mechanics, it leads to the complicated big bang cosmology and inflation scenarios. In this theme, it is important to know how to quantize string and D-brane actions \cite{wi,kw}. As a matter of fact, this is possible using Dirac's point of view, since these actions include primary second class constraints and actually are not gauge invariant. 

As the pioneer who proposed the correlation between gauge theories and constrained systems, Dirac classified constraints into first and second class ones. He also implied that the existence of second class constraints is due to the presence of some nonphysical degrees of freedom, which destroys the gauge invariance \cite{1}. Hence, these extra degrees of freedom must be omitted or changed to physical ones to enhance gauge symmetries of the model \cite{torus}. In this regards, there are several method available to convert second class systems into first class ones \cite{1,torus,2}.

Due to the presence of gauge degrees of freedom, quantization of a first class system is straightforward, but dealing with second class systems arise some difficulties. Quantizing a system which includes second class constraints is not possible due to the presence of extra coordinates in the primary phase space. Dirac solved this incompatibility by converting Poisson brackets into another kind of brackets, which are known as Dirac brackets in respect of him, today. Although his approach is serviceable in some cases, dealing with Dirac brackets is not generally an easy task. Some difficulties such as factor ordering and inverting Poisson brackets matrix of constraints may cause Dirac's approach not practical. \cite{11,14,15,20}.

A question might have been raised that why we do not vanish second class constraints directly? One should say that it is not always possible to vanish second class constraints, while it may omit some of the dynamics of the system automatically. It is more wise to convert second class constraints into first class ones, which leaves the dynamics of the system intact. This has been shown in some models that the embedding process do not change the dynamics of the system \cite{torus, anyon, NC, Mardani}.

To convert a second class system to a first class one, there have been existed several methods such as gauge unfixing approach (GU) \cite{gu1,gu2}, BFT method \cite{16,17,18,19}, and the symplectic formalism \cite{NC,FJ1, FJ2, FJ3}.

The BFT method has been proposed to convert a gauge non-invariant system to a gauge invariant one. This method is based on embedding procedure, i.e. to convert second class constraints into first class ones as gauge symmetries, one should increase phase space of the primary model and redefine constrained equations in the extended phase space.

Depending on the structure of second class constraints, the extended phase space have different forms. Sometimes, the procedure of BFT embedding is complicated, even impossible to do. It has been shown that if the Poisson bracket matrix of second class constraints has constant components, extension of the phase space is straightforward and added correction terms to Hamiltonian will be limited in number \cite{20}.

In this research, according to constraints' algebra which leads to the constant Poisson bracket matrix, we use finite BFT approach to convert second class constraints to first class one, in order to obtain the gauged version of Open 2-brane system. 

The interesting and almost problematic point in the study of Open string problem in the presence of background magnetic field is that after imposing mixed boundary conditions on equations of motion, the fields' Poisson brackets are not vanished at boundaries and they are equal to a constant value which is proportional to the present background field \cite{3,4,5,6,7}. This problem also leads to a contradiction when we study secondary constraints.

The idea of considering boundary conditions as constraints of a physical system was proposed \cite{8,9,10} by the late 1990s. In fact, boundary conditions in the background magnetic field are first order equations of time variable, whereas the Euler-Lagrange equation is a second order one, which means that the acceleration determines the dynamics of the motion. Hence, boundary conditions in Lagrangian equations are called acceleration-free equations \cite{4,5,6,7,10}. So, these objects do not play any role in the dynamics of the system. Instead, they set up some identities in the phase space which satisfy Dirac's constrained conditions. Also, the difference between constraints, obtained from boundary conditions equations and other constraints could be evident in Hamiltonian approach \cite{10}.

By the idea of boundary conditions as Dirac conditions, we study the constrained structure of bosonic p-brane systems \cite{DF,12,13}. First of all, we simplify the action of this theory, based on existed symmetries of the model. Then, constrained structure of the model with $p=2$, i.e. bosonic open p-brane system, is obtained and the consistency condition of constraints are checked in order to obtain secondary constraints. Considering Poisson brackets of constraints, we prove the second class nature of them.

\section{Action of Open p-brane}
D-branes are particular types of branes. They are places in the space where strings are located. $D_{p} $ brane is stretched out in P spatial dimensions which sweeps out a $(P+1)$ dimensional world volume \cite{21,22}. The action studied in this paper is the bosonic part of a classical action for an Open p-brane ending on a $D_{q} $-brane \cite{13,23}. In this procedure, we avoid probing other aspects such as terms related to fermionic parts of the action as much as possible. Also, in order to decrease the generality of our research, we only consider the aspects of action which impose mixed boundary conditions.
\begin{equation}
S=-\frac{1}{4\pi \acute{\alpha}}\int_{\sum} d^{p+1}\zeta \sqrt{-h}[G_{\mu \nu} h^{\alpha\beta}\frac{\partial X^{\mu}}{\partial \zeta^{\alpha}}\frac{\partial X_{\mu}}{\partial \zeta^{\beta}}+(p-1)].
\label{1}
\end{equation}
The values of $\mu$ varies from zero to D. Also, $\alpha,\beta =\{ 0,1,...,p\}$ and $ (p-1) $ is the cosmological term.

This model has a large surface of global symmetries, which are determined by the world volume and the target space. To study a simpler model, we consider the dynamics of fields in the flat target space.
\begin{equation}
G_{\mu\nu}=\eta_{\mu\nu}.
\label{2}
\end{equation}

Also, we let $\acute{\alpha}$ to be a constant numerical parameter. Moreover, the world volume has diffeomorphisms and scaling invariance. These symmetries can be fixed by determining a metric on it. We consider this metric as  $ h _{\alpha\beta}= (-,+,+,...,+)$ .

The canonical momentum and the Hamiltonian are defined as follows,
\begin{eqnarray}\label{4}
&& \Pi_{\mu} = \partial_{\tau} X^{\mu}   ,  \\ \nonumber 
&& H=\frac{1}{2}\int \frac{d^{p}\zeta}{(2\pi)^{p}}[(\Pi_{i})^{2}+(\Pi_{a})^{2}+\displaystyle\sum^{p}_{k=1}(\partial_{k} X^{k})^{2}+\displaystyle\sum^{p}_{k=1}(\partial_{k} X^{a})^{2}+(p-1)],
\end{eqnarray}
where $\partial_{k} = \frac{\partial}{\partial \sigma^{k}}$,\quad $i=0,1,...,q$ , \quad$a=q+1,...,D$.

It is evident that, in directions orthogonal to D branes,the  boundary condition is the form of Dirichlet boundary condition  and toward D-brane is the form of Neumann boundary condition. Assuming $q\geq p$, i.e. the dimension of D-brane is less than or equal to the space dimension, we can only have Neumann boundary condition.
Dynamic of boundary points with Neumann boundary conditions are as follows,
\begin{eqnarray}\label{4-1}
\partial_{k} X^{i}(0)=\partial_{k} X^{i}(\pi) = 0. 
\end{eqnarray}
Obviously these equations are at most first class derivatives respect to the time and are called acceleration-free equations. As we mentioned before, such equations do not play any roles in present object dynamics of the theory and they only limit available phase spaces, thus they satisfy Dirac constrained condition.

\section{Constraint structure of Open 2-brane System}
It is important to know how to quantize String and D-brane action.
The alternative method is solving all of the constraints of the system at first. 
So, In this section  and as an example, we analyse Open 2-brane system. As we know, a 2-brane is surrounded by a three dimensions hyper surface.
The Neumann boundary conditions are imposed at the end of Open 2-brane \cite{13},
\begin{equation}\label{3}
\partial_{1}X^{i}(0)=\partial_{1}X^{i}(\pi)  \ \ \ , \ \ \  \partial_{2}X^{i}(0)=\partial_{2}X^{i}(\pi).\\
\end{equation}
The boundary conditions can be rewritten in terms of normal 2-brane modes.
\begin{eqnarray}\label{4-2}
&&^{(1)}\phi^{i(0)}_{m}= \displaystyle\sum _{n}  n X^{i}_{nm}\approx 0 \ \ , \ \  ^{(1)}\bar{\phi}^{i(0)}_{m}=\displaystyle\sum _{n}(-1)^{n} n X^{i}_{nm}\approx 0\nonumber \\
&&^{(2)}\phi^{i(0)}_{n}=\displaystyle\sum _{m} m X^{i}_{nm}\approx 0 \ \ \ , \ \ ^{(2)}\bar{\phi}^{i(0)}_{n}=\displaystyle\sum _{m}(-1)^{m} m X^{i}_{nm}\approx 0.
\end{eqnarray}
where, $\approx$ means weak equality.

The canonical Hamiltonian can be written as,
\begin{eqnarray}
& H=\frac{1}{2}\displaystyle\sum_{n}\displaystyle\sum_{m}\eta_{ij}[P_{nm}^{i}P_{(-n)(-m)}^{j}+(n^{2}+m^{2})X_{nm}^{i}X_{(-n)(-m)}^{j}]  \nonumber \\
&\qquad +\frac{1}{2} \displaystyle\sum_{n}\displaystyle \sum_{m}\eta_{ab}[P_{nm}^{a}P_{(-n)(-m)}^{b}+(n^{2}+m^{2})X_{nm}^{a}X_{(-n)(-m)}^{b}].
\end{eqnarray}
Hence, primary constraints must be hold in all the time, we set the complete time derivative of first class constraints equal to zero. in this manner, the consistency procedure gives the set of secondary constraint \cite{13}.
\begin{eqnarray}\label{4-3}
&&^{(1)}\Psi^{i(0)}_{m}=\lbrace H, ^{(1)}\phi^{i(0)}_{m}\rbrace_{PB}=\displaystyle\sum _{n}  n P^{i}_{nm} \approx 0 \nonumber \\
&&^{(1)}\bar{\Psi}^{i(0)}_{m}=\lbrace H, ^{(1)}\bar{\phi}^{i(0)}\rbrace_{PB}=\displaystyle\sum _{n} (-1)^{n} n P^{i}_{nm} \approx 0\nonumber \\
&&^{(2)}\Psi^{i(0)}_{m}=\lbrace H, ^{(2)}\phi^{i(0)}_{m}\rbrace_{PB}=\displaystyle\sum _{n}  m P^{i}_{nm} \approx 0\nonumber \\
&&^{(2)}\bar{\psi}^{i(0)}=\lbrace H, ^{(2)}\bar{\phi}^{i(0)}\rbrace_{PB}=\displaystyle\sum _{n} (-1)^{m} m P^{i}_{nm} \approx 0.
\end{eqnarray}

By further imposing the  time consistency of the secondary constraints, new constraints will be obtained as follows,
\begin{eqnarray}\label{4-4}
&&^{(1)}\phi^{i(1)}_{m}=\lbrace H, ^{(1)}\Psi^{i(0)}_{m}\rbrace_{PB}=-\displaystyle\sum _{n}  n(n^{2} +m^{2})X^{i}_{nm}\approx 0 \nonumber \\
&&^{(1)}\bar{\phi}^{i(1)}_{m}=\lbrace H, ^{(1)}\bar{\Psi}^{i(0)}_{m}\rbrace_{PB}=-\displaystyle\sum _{n}  (-1)^{n}n(n^{2} +m^{2})X^{i}_{nm}\approx 0\nonumber \\
&&^{(2)}\phi^{i(1)}_{n}=\lbrace H, ^{(2)}\Psi^{i(0)}_{m}\rbrace_{PB}=-\displaystyle\sum _{m}  m(n^{2} +m^{2})X^{i}_{nm}\approx 0\nonumber \\
&&^{(2)}\bar{\phi}^{i(1)}_{m}=\lbrace H, ^{(2)}\bar{\phi}^{i(0)}\rbrace_{PB}=-\displaystyle\sum _{n} (-1)^{m} n(n^{2} +m^{2})X^{i}_{nm}\approx 0.
\end{eqnarray}
 Continuing this procedure, we will obtain the complete chain structure of constraints of the model \cite{torus}.
Hence, we will have,
 \begin{eqnarray*}
^{(1)}\phi^{i(k)}_{m}&=& (-1)^{k}\displaystyle\sum _{n}  n(n^{2} +m^{2})^{k}X^{i}_{nm}\approx 0\\
^{(1)}\bar{\phi}^{i(k)}_{m}&=&(-1)^{k}\displaystyle\sum _{n}  (-1)^{n}n(n^{2} +m^{2})^{k}X^{i}_{nm}\approx 0\\
^{(2)}\phi^{i(k)}_{n}&=& (-1)^{k}\displaystyle\sum _{m}  m(n^{2} +m^{2})^{k}X^{i}_{nm}\approx 0\\
^{(2)}\bar{\phi}^{i(k)}_{n}&=&(-1)^{k}\displaystyle\sum _{m}  (-1)^{m}n(n^{2} +m^{2})^{k}X^{i}_{nm}\approx 0\\
^{(1)}\Psi^{i(k)}_{m}&=& (-1)^{k}\displaystyle\sum _{n}  n(n^{2} +m^{2})^{k}P^{i}_{nm}\approx 0\\
^{(1)}\bar{\Psi}^{i(k)}_{m}&=&(-1)^{k}\displaystyle\sum _{n}  (-1)^{n}n(n^{2} +m^{2})^{k}P^{i}_{nm}\approx 0\\
^{(2)}\Psi^{i(k)}_{n}&=& (-1)^{k}\displaystyle\sum _{m}  m(n^{2} +m^{2})^{k}P^{i}_{nm}\approx 0\\
^{(2)}\bar{\Psi}^{i(k)}_{n}&=&(-1)^{k}\displaystyle\sum _{m}  (-1)^{m}n(n^{2} +m^{2})^{k}P^{i}_{nm}\approx 0.
\end{eqnarray*}
Summing over $ k $s and noting the fact that the result must be equal to zero, these constraints make a close set. Hence, we may find the following sets of equations of constraints which obviously are independent from $k$.

\begin{align}\label{8}
&&^{(1)}\chi^{i}_{nm} =X_{nm}^{i}-X_{(-n)m}^{i}\approx 0  \ \ \ \ , \ \ \  ^{(2)}\chi^{i}_{nm} =X_{nm}^{i}-X_{n(-m)}^{i}\approx 0
\nonumber \\
&&^{(1)}\varphi^{i}_{nm} =P_{nm}^{i}-P_{(-n)m}^{i}\approx 0  \ \ \ , \ \ \ ^{(2)}\varphi^{i}_{nm} =P_{nm}^{i}-P_{n(-m)}^{i}\approx 0.
\end{align}

\section{Gauging the Non-invariant Open 2-brane Model}
The emergence of second class constraints in the model is due to the broken gauge symmetry.
As we mentioned before, the gauging process will be done via the BFT method.
This method is based on extending the phase space of the original model by introducing some auxiliary variables.

considering a pure second class system which its dynamics is defined by the Hamiltonian $H^{(0)}$ in the phase space with coordinates $(p,q)$ and a set of second class constraints $\Theta_{\alpha}$  in which $ \alpha =\{ 1,\ldots ,m\} $  , we will have the Poisson bracket matrix as the following invertible matrix.
\begin{equation}
\Delta_{\alpha\beta}=\{\Theta_{\alpha},\Theta_{\beta}\}
\end{equation}
For our model, the gauge structure constructing Poisson bracket matrix of second class constraint is constant matrix as follows,
\begin{equation}
\Delta=\left(
         \begin{array}{cccc}
         0 & 0 & 2\eta^{ij} & \eta^{ij} \\
         0 & 0 &\eta^{ij} & 2\eta^{ij} \\
         -2\eta^{ij}  & -\eta^{ij}  & 0 & 0 \\
         -\eta^{ij}  & -2\eta^{ij} & 0 & 0 \\
        \end{array}
        \right)
\end{equation}

Now, we enlarge the phase space by the variable $\xi^{\alpha}$ which satisfies following Algebra \cite{16,17,18,19,20}.
\begin{eqnarray}\label{4-5}
&& \lbrace q_{i},p_{i}\rbrace=\delta_{ij}  \ \ \ \ ,\ \ \  \lbrace q_{i},q_{j}\rbrace=\lbrace p_{i},p_{j}\rbrace=0\\
&& \lbrace \xi_{\alpha},q_{i}\rbrace=\lbrace \xi_{\alpha},p_{i}\rbrace=0 \ \ \ ,\ \ \  \lbrace \xi_{\alpha},\xi_{\beta}\rbrace=\omega_{\alpha\beta},
\end{eqnarray}
where $\omega^{\alpha\beta}$ is an antisymmetric invertible matrix. The first class constraints in the extended phase space $(p,q)\oplus\xi$ are defined by,
\begin{equation}\label{4-6}
\tau_{\alpha}=\tau_{\alpha}(p,q,\xi) \ \ \ \ \ \ \ \ \alpha =1,2,\ldots,m,
\end{equation}
which satisfy the following boundary condition.
\begin{equation}\label{4-7}
\tau^{(0)}_{\alpha}(p,q,0)=\Theta_{\alpha}(p,q).\\
\end{equation}

For special cases where the matrix $\Delta_{\alpha\beta}$  is sympletic or constant, a systematic method is invented which proposes following choices \cite{20},
\begin{equation}\label{8-2}
\omega = -\Delta \ \ \ \ \ , \ \ \ \ \ \Omega = 1.
\end{equation}

These choices lead to the following finite order embedding formalism for constraints,
\begin{equation}\label{1-1}
\tau_{\alpha}=\tau_{\alpha}^{(0)}+\xi^{\alpha}.
\end{equation}

Another important point is that, in order to enlarge the phase space in String theory we should introduce tensor fields, in contrast with other theories that scaler and vector fields must be added. So the new set of constraints will be,
\begin{align}\label{8-3}
& ^{(1)}\tau^{i}_{nm}= ^{(1)}\chi^{i}_{nm} + ^{(1)}\xi^{i}_{nm} \ \ \ \ , \ \ \  \ ^{(2)}\tau^{i}_{nm}= ^{(2)}\chi^{i}_{nm} + ^{(2)}\xi^{i}_{nm}
 \nonumber \\
& ^{(3)}\tau^{i}_{nm}= ^{(1)}\varphi^{i}_{nm} + ^{(3)}\xi^{i}_{nm}\ \ \ \ , \ \ \  ^{(4)}\tau^{i}_{nm}= ^{(2)}\varphi^{i}_{nm} + ^{(4)}\xi^{i}_{nm}.
\end{align}

The generators of Hamiltonian correction terms \cite{16,17,18,19,20} will be,
\begin{align}
& G^{ (0)}_{1}=\{X_{nm}^{i}-X_{(-n)m}^{i},H^{(0)}\} = 2(P_{(-n)(-m)}^{i}-P_{n(-m)}^{i})\nonumber \\
& G^{ (0)}_{2}=\{X_{nm}^{i}-X_{n(-m)}^{i},H^{(0)}\} = 2(P_{(-n)(-m)}^{i}-P_{(-n)m}^{i})\nonumber \\
& G^{(0)}_{3}= \{P_{nm}^{i}-P_{(-n)m}^{i},H^{(0)}\} = 2(n^{2}+m^{2})(X_{(-n)(-m)}^{i}-X_{n(-m)}^{i})\nonumber \\
& G^{ (0)}_{4}=\{P_{nm}^{i}-P_{n(-m)}^{i},H^{(0)}\} = 2(n^{2}+m^{2})(X_{(-n)(-m)}^{i}-X_{(-n)m}^{i}).
\end{align}

Hence, first order Hamiltonian is obtained as,
\begin{align}
 &\tilde{H}^{(1)}=-2\eta^{ij}[(P_{(-n)(-m)}^{i}-P_{n(-m)}^{i})(2 ^{(3)}\xi^{i}_{nm}+ ^{(4)}\xi^{i}_{nm})\nonumber \\
&\quad \qquad -(P_{(-n)(-m)}^{i}+P_{(-n)m}^{i})(2 ^{(4)}\xi^{i}_{nm}+ ^{(3)}\xi^{i}_{nm})\nonumber  \\
& \quad \qquad +2(n^{2}+m^{2})(X_{(-n)(-m)}^{i}-X_{n(-m)}^{i})(2 ^{(1)}\xi^{i}_{nm}+ ^{(2)}\xi^{i}_{nm})\nonumber \\
& \quad \qquad +(X_{(-n)(-m)}^{i}-X_{(-n)m}^{i})(^{(1)}\xi^{i}_{nm}+ 2 ^{(2)}\xi^{i}_{nm})],
\end{align}
and the generators in the second correction term of Hamiltonian is investigated as,
\begin{eqnarray}
&& G^{ (1)}_{1}=\{X_{nm}^{i}-X_{(-n)m}^{i},\tilde H^{(1)}\} \\
&& \qquad= -6 ^{(3)}\xi^{i}_{(-n)(-m)}+6 ^{(3)}\xi^{i}_{n(-m)}+ 2 ^{(3)}\xi^{i}_{(-n)m}-2 ^{(3)}\xi^{i}_{nm}\nonumber \\
&& \qquad +4 ^{(4)}\xi^{i}_{(-n)m}-2 ^{(4)}\xi^{i}_{nm} \nonumber \\
\nonumber \\
&&G^{ (1)}_{2}=\{X_{nm}^{i}-X_{n(-m)}^{i},\tilde H^{(1)}\}   \\
&&\qquad= -6 ^{(3)}\xi^{i}_{(-n)(-m)}+4 ^{(3)}\xi^{i}_{n(-m)}+2 ^{(3)}\xi^{i}_{(-n)m}+2 ^{(3)}\xi^{i}_{nm} \nonumber \\  
&&\qquad \quad +3 ^{(4)}\xi^{i}_{(-n)(-m)}-2 ^{(4)}\xi^{i}_{n(-m)}-9 ^{(4)}\xi^{i}_{(-n)m}+2 ^{(4)}\xi^{i}_{nm}\nonumber
\end{eqnarray}
\begin{eqnarray}
&&G^{ (1)}_{3}=\{P_{nm}^{i}-P_{(-n)m}^{i},\tilde H^{(1)}\}\\ 
&&\qquad= (n^{2}+m^{2})[10 ^{(1)}\xi^{i}_{(-n)(-m)} - 6 ^{(1)}\xi^{i}_{n(-m)}-2 ^{(1)}\xi^{i}_{(-n)m}+2 ^{(1)}\xi^{i}_{nm}\nonumber \\ 
&&\qquad \quad +8 ^{(2)}\xi^{i}_{(-n)(-m)}- 8 ^{(2)}\xi^{i}_{n(-m)}-4 ^{(2)}\xi^{i}_{(-n)m}+2 ^{(2)}\xi^{i}_{nm}]\nonumber \\
\nonumber \\
&&G^{ (1)}_{4}=\{P_{nm}^{i}-P_{n(-m)}^{i},\tilde H^{(1)}\} \\
&&\qquad= (n^{2}+m^{2})[8 ^{(1)}\xi^{i}_{(-n)(-m)}-8 ^{(1)}\xi^{i}_{(-n)m}-4 ^{(1)}\xi^{i}_{n(-m)}\nonumber \\
&&\qquad \quad +10 ^{(2)}\xi^{i}_{(-n)(-m)}-2 ^{(2)}\xi^{i}_{(-n)(-m)}-10 ^{(2)}\xi^{i}_{(-n)m}-2 ^{(2)}\xi^{i}_{nm}].\nonumber
\end{eqnarray}

Hence, the second order correction of Hamiltonian is obtained as,
\begin{eqnarray}
&&\tilde{H}^{(2)}=\frac{-1}{2}\eta^{ij}[(2^{(3)}\xi^{i}_{nm}+^{(4)}\xi^{i}_{nm})G_{1}^{(1)}-(2^{(1)}\xi^{i}_{nm}+^{(2)} \xi^{i}_{nm})G_{3}^{(1)}\nonumber \\
&&\qquad \quad +(2^{(4)}\xi^{i}_{nm}+^{(3)}\xi^{i}_{nm})G_{2}^{(1)}-(^{(1)}\xi^{i}_{nm}+2^{(2)}\xi^{i}_{nm})G_{2}^{(1)}].
\end{eqnarray}

It is easy to see that $G^{(n)}_{\alpha}$ vanished for $n\geq2$. This leads to the truncated series for Hamiltonian for $n\geq2$.
\begin{equation}
H_{T}=H^{(0)}+\tilde{H}^{(1)}+\tilde{H}^{(2)}.
\end{equation}

Thus, by introducing some tensor fields, first class constraints of second ones would be obtained and the Hamiltonian investigated in such away will be fully gauged in the extended phase space.
\section{Conclusion}
String theory is an extremely good theory to describe the world. So, quantizing String and D-brane action is an interesting topic to work on. A appropriate approach to do such an investigation is the formalism of constrained systems.

 In this survey, first, we extracted the constrained structure of the Open 2-brane System in  the flat space. Because the boundary conditions are usually relations between fields and their derivatives, we considered the Neumann boundary conditions as Dirac constraints which are not consequences of a singular Lagrangian. In other words, the momenta are not independent functions of velocities which is a new feature in the context of constrained systems. Finally, we got four second class constraints. As we know, first class constraints are the generators of gauge transformations, while
 the advent of second class constraints restricts the system to a smaller sub-manifold of the phase space in which a Poisson structure can be recognized. This means that gauge symmetries are broken.
 
 We attended the quantization of theory but quantization of second class system is non-trivial and is more difficult in comparison with first class one. Difficulties such as factor ordering problem or non-invertible nature of some Poisson brackets matrices of constraint may arise dealing with second class systems. Also, the construction of a BRST charge is only possible for first class systems. More importantly, the usual quantization method like canonical quantization and path integral approach is only used for  first class constrained systems. Since, physical theories tend to be gauged theories, and the presence of second class constraints is against with this assumption, we convert the Non-invarient Open 2-brane System to a first class one by means of BFT method .
 
 As we mentioned,  for special cases when the  $ \Delta-$ matrix has constant elements, the Finite Order BFT embedding can be applied. In the last section we found correction terms for constraints and Hamiltonian in the extended phase space. All in all, one can easily check that first class constraints and Hamiltonian of Open 2-brane System represents a gauge invariant theory.

\section*{Acknowledgement}
We wish to appreciate the University of Kashan for the Grant No. 65500.3\\

\end{document}